\documentclass[preprint,5p,times, twocolumn]{elsarticle}

\usepackage{amssymb}

 \usepackage{lineno}

\usepackage{amssymb,amsmath}
\usepackage{graphicx,float}
\usepackage{hyperref}
\usepackage{diagramas} 
\usepackage{color}

\newcommand{\beq}{\begin{equation}}
\newcommand{\eeq}{\end{equation}}
\newcommand{\bea}{\begin{eqnarray}} 
\newcommand{\eea}{\end{eqnarray}}

\def\({\left(}
\def\){\right)}

\journal{Physics Letters A}

\begin{document}

\begin{frontmatter}



 \title{Further investigation of mass dimension one fermionic duals}

\author[affiliation]{J. M. Hoff da Silva}
\ead{julio.hoff@unesp.br}
\author[affiliation]{R. T. Cavalcanti\corref{cor1}}
\ead{rogerio.cavalcanti@unesp.br}
\address[affiliation]{Departamento de F\'isica e Qu\'imica, Universidade Estadual Paulista (UNESP),\\
 Guaratinguet\'a-SP-Brazil}
 

\author{}

\address{}

\begin{abstract}
In this paper we proceed into the next step of formalization of a consistent dual theory for mass dimension one spinors. This task is developed approaching the two different and complementary aspects of such duals, clarifying its algebraic structure and the so called $\tau-$deformation. The former regards the mathematical equivalence of  the recent proposed Lorentz preserving dual with the duals of algebraic spinors, from Clifford algebras, showing the consistency and generality of the new dual. Moreover, by revealing its automorphism structure, the hole of the $\tau-$deformation and contrasting the action group orbits with other Lorentz breaking scenarios, we argue that the new mass dimension one dual theory is placed over solid and consistent basis.

\end{abstract}


\end{frontmatter}



\section{Introduction}

The intriguing and rich theory behind mass dimension one dark spinor and spinorial structure has revealed a promising field of research pointing towards one of biggest challenges of the XXI century physics, namely the dark matter problem. In spite of the relevance of mass dimension one spinors (Elko) \cite{Ahluwalia:2004sz,Ahluwalia:2004ab} by itself as candidate to describe dark matter, it has raised important questions concerning, for example, exotic dark structure \cite{daRocha:2011yr}, algebraic classification \cite{Lounesto:2001zz,Cavalcanti:2014wia,HoffdaSilva:2017waf,Fabbri:2017lvu}, representatives of new classes \cite{Lee:2018bfv,daRocha:2013qhu,daRocha:2016bil}, magnetic monopoles \cite{deOliveira:2013hda} as well as torsional gravity \cite{Fabbri:2014foa}, Elko \cite{Vaz:2017fac} and general spinors dynamics \cite{Fabbri:2017pwp}.
In addition, as evinced in Ref. \cite{Ahluwalia:2016rwl}, the freedom of choice of spinors dual is another central issue motivated by the mass dimension one spinors theory. 

The quantum field theory literature usually takes the Dirac dual as the standard one, with no suspicion or need of alternative dual structures as being potentially interesting. However, the development of the theory of dark spinors naturally requires a different dual \cite{Ahluwalia:2004sz, Ahluwalia:2004ab}. The initial proposal suffered the lack of Lorentz invariance, demanding a careful investigation of how general a physics motivated dual spinor candidate can be. An answer to this issue was the emergence of the new mass dimension one dual, proposed in \cite{Ahluwalia:2016rwl},  following a twofold procedure, namely the new dual definition and its $\tau-$deformation. Both necessary to perform a full Lorentz covariant (and local) quantum field theory. Our aim here is to elucidate these two aspects.

The algebraic theory of spinor duals makes use of the rich and well known structure of Clifford algebras to specify all possible duals for arbitrary algebras of any dimension and space(time) signature \cite{Benn:1987fw}.  Here emerges the confrontation of two very different, 	
although algebraically equivalent, dual definitions. By comparing the new dual raised in the dark spinors theory to the one from the algebraic definition of spinors, we show that the former is actually a slightly particular, yet interesting, case of the later. We proceed by going further on the mass dimension one duals, evincing a hidden automorphism structure behind the so called $\tau$-deformation \cite{Rogerio:2016mxi}. It makes the choice made in \cite{Ahluwalia:2016rwl} (see Sec. \ref{prel}) an execllent candidate for the right dual ensuring the Lorentz invariance of the theory. The hole of the $\tau$-deformation on Lorentz invariance is also discussed by comparing its effects with other scenarios of symmetry breaking and orbits of a set of spin sums under the action of very special relativity groups \cite{Cohen:2006ky}. Within the context previously exposed, we approach both parts of the new dual formulation, thereby reinforcing its consistence and generality. 

The present paper is organized as follows: In Sec. \ref{prel} we briefly introduce the problem of mass dimension one duals and the solution proposed in \cite{Ahluwalia:2016rwl} in order to keep the theory Lorentz invariant. It is followed, in Sec. \ref{algebraic}, by a brief exposition of Clifford algebras and the definition of spinors as minimal ideals of such algebras, as well as the algebraic structure of general duals, in order to contrast it with the one developed under the mass dimension one spinors theory. In Sec. \ref{auto} we expose the automorphism structure of the $\tau$-deformed dual as well as its importance in Lorentz symmetry preserving.  We finish the paper by discussing the results and its importance on the task of constructing a complete and consistent theory of mass dimension one spinors.   

\section{Preliminaries}\label{prel}

The idea that the usual Dirac dual cannot be applied to every spinor is sharp enough to force the development of an accurate criteria in the formalization of spinor duals, recovering the usual case in an appropriate limit. Most of the formalization concerning spinor duals can be found in Section 4 of Ref. \cite{Ahluwalia:2016rwl}. Here we shall pinpoint the main steps of finding a generalized spinor dual and, then, move forward in the formalization process.  

A given spinor field, $\psi$, carrying a $(1/2,0)\oplus(0,1/2)$ representation is supposed to have a dual given by 
\begin{equation}
\tilde{\psi_s}(p^\mu)=[\Delta \psi_s(p^\mu)]^{\dagger}\eta, \label{1}
\end{equation} for which judicious (physical) constraints of the theory shall reveal the $\Delta$ operator and the $\eta$ matrix as well. In the above equation, $s$ stands for the spinor helicity structure (the Elko's are formally constructed to have a double helicity structure given rise to four spinors, each one labeled by the index $s$). Lorentz invariance of the product $\tilde{\psi}_s\psi_s$ implies (up to an irrelevant constant) $\eta=\gamma_0$. For the $\Delta$ operator it is allowed to change the helicity structure of the spinor upon which it acts. It can be shown that the helicity structure being unchanged and having at hand a typical Dirac spinor, then $\Delta=\mathbb{I}$ and we have the usual Dirac dual. On the other hand, if the helicity structure is changed, then one may be pushed to a more elaborated operator \cite{Ahluwalia:2016rwl} given by $\Delta=m^{-1}G(\phi)\gamma_\mu p^{\mu}$, where $p^\mu$ is the particle momenta, $m$ its rest mass and  
\begin{eqnarray}\label{gphi}
G(\phi)= \left(\begin{array}{cccc}
0 & 0 & 0 & -ie^{-i\phi} \\ 
0 & 0 & ie^{i\phi} & 0 \\ 
0 & -ie^{-i\phi} & 0 & 0 \\ 
ie^{i\phi} & 0 & 0 & 0
\end{array}  \right),
\end{eqnarray} with $\phi$ being the polar angle in the momentum parameterization $p^{\mu}=(E,p\sin\theta\cos\phi,p\sin\theta\sin\phi,p\cos\theta)$.

The relevant spinorial structure for mass dimension one operators is attained by requiring neutrality with respect to the charge conjugation operator, i. e., $C\lambda_h^{S/A}=\pm \lambda_h^{S/A}$, given rise to the self-conjugated ($\lambda_h^S$) and the anti-self-conjugated spinor ($\lambda_h^A$). The resulting spin sums, calculated after the formal spinor structure be settled, reads 
\begin{eqnarray}
\sum_h \lambda^{S/A}_h(p^\mu)\tilde{\lambda}^{S/A}_h(p^\mu)=\pm m[\mathbb{I}\pm G(\phi)],\label{4}
\end{eqnarray} explicitly breaking Lorentz invariance (see Refs. \cite{Lee:2015jpa, Lee:2015sqj} for a thorough analysis of mass dimension one fermions interactions and symetries). The systematic study of the symmetries encoded in $G(\phi)$ reveals that it is, in fact, invariant (or form covariant) under the transformations of $HOM(2)$ ($SIM(2)$) \cite{Ahluwalia:2010zn, Ahluwalia:2008xi}, both subgroups of the Lorentz group obtained by removing the discrete symmetry operators $P$ and $T$ and rearranging the remain generators. These subgroups entails the underlining symmetry of the so-called Very Special Relativity (VSR) \cite{Cohen:2006ky}. The attempt to restore, at the formal level, the Lorentz symmetry of the spin sums went over the following combined procedure. First, a redefinition of the dual was implemented such as \cite{Ahluwalia:2016rwl}
\begin{eqnarray}\label{A}
\stackrel{\neg}{\lambda}\,\!\!^S_h(p^\mu)= \tilde{\lambda}^{S}_h(p^\mu)\mathcal{A}, \\ \label{B}
\stackrel{\neg}{\lambda}\,\!\!^A_h(p^\mu)= \tilde{\lambda}^{A}_h(p^\mu)\mathcal{B},
\end{eqnarray}
where the operators $\mathcal{A}$ and $\mathcal{B}$ must obey the important and nontrivial properties:
\begin{eqnarray}\label{7}
\mathcal{A}\lambda^{S}_h(p^\mu) = \lambda^{S}_h(p^\mu), \quad\quad \mathcal{B}\lambda^{A}_h(p^\mu)=\lambda^{A}_h(p^\mu),
\end{eqnarray} along with
\begin{eqnarray}\label{8}
\stackrel{\neg}{\lambda}\,\!\!^S_h(p^\mu)\mathcal{A}\lambda^{A}_h(p^\mu) = 0, \quad\quad \stackrel{\neg}{\lambda}\,\!\!^A_h(p^\mu)\mathcal{B}\lambda^{S}_h(p^\mu) = 0,
\end{eqnarray} in order to the orthonormality relations remain unchanged, a surely indispensable issue in a spinor based theoretical formulation. In Sec. \ref{auto} we shall take advantage of the formal properties of $\mathcal{A}$ and $\mathcal{B}$ and Sec. \ref{algebraic} is devoted to evince the Clifford algebra procedure behind such a redefinition. Nevertheless, before to investigate these points, we would like to finalize this short preliminary section by showing that the new spin sums, calculated with the aid of the new duals, amounts to 
\begin{eqnarray}\label{spinsumab}
\sum_h \lambda^{S}_h(p^\mu)\stackrel{\neg}{\lambda}\,\!\!^S_h(p^\mu)=m[\mathbb{I}+G(\phi)]\mathcal{A}, \\
\sum_h \lambda^{A}_h(p^\mu)\stackrel{\neg}{\lambda}\,\!\!^A_h(p^\mu)=-m[\mathbb{I}-G(\phi)]\mathcal{B}
\end{eqnarray} and the Lorentz invariance requires also the deformation of the $(\mathbb{I}\pm G(\phi))$ operator in order to allow the existence of an inverse. In fact, since $det(\mathbb{I}+G(\phi))=0$, the following stratagem was used: to deform the initial operator by means of a real parameter $\tau$ in such a way that the spin sums become proportional to $(\mathbb{I}\pm \tau G(\phi))$ and the limit $\tau\rightarrow 1$ must be taken at the final of the calculations. This procedure, implemented in Ref. \cite{Lee:2014opa}, have formal validity, as emphasized in Ref. \cite{Rogerio:2016mxi}. In Sec. \ref{auto} we show that this $\tau-$deformation, due to the very nature of $\mathcal{A}$ and $\mathcal{B}$ operators, is indeed a profound necessity. Before that, in the next section we shall thoroughly investigate the formal and general structure of the dual in Eq. \eqref{1}, approaching hence the two main aspects of the new dual theory.

\section{The algebraic view of the ``new'' duals}\label{algebraic}

It is a well known fact that the formal aspects of spinors has been highly enlightened by  thorough studies of such objects defined over the structure of Clifford algebras.  Our aim in this section is to revisit the algebraic definition of spinors and analyze the general form of the dual of Eq. \eqref{1} from this point of view. We start by introducing the basic structure needed to define the algebraic spinors space of a general Clifford algebra and inner products on such spaces, as well as in its complexified version. The comparison of the complexified algebraic dual spinors with the one of Eq. \eqref{1} justifies the quotation marks at the section title. As we are interested in general and algebraic properties of spinors duals, along this paper we shall drop the helicity structure index as well as the spacetime point dependence of spinors and operators, except when strictly necessary.

Clifford algebras are rich algebraic structures, defined over quadratic spaces, with application ranging from theoretical physics to engineering and computer science (See Ref. \cite{Hitzer2013} for a review with a wide range of applications of Clifford algebras). Its multivectorial graded structure encompass the tensorial and exterior algebra as well as the quaternions ring. The Spin groups, isomorphic to classical Lie groups in lower dimensions, associated to its clear geometric view, provide a general and elegant way to define rotations on arbitrary quadratic spaces\footnote{For comprehensive and accessible books on Clifford algebras and its applications in physics see, for example, Refs. \cite{Lounesto:2001zz} and \cite{Vaz:2016qyw}.}.

Given a real vector space equipped with a symmetric quadratic form $g$ of signature $(p,q)$, denoted by $\mathbb{R}^{p,q}$, its associated Clifford algebra is defined as follows:\footnote{{As it does not involve additional complexity, we start discussing the Clifford algebra associated to an arbitrary real space with signature $(p,q)$. Then we particularize to $\mathbb{R}^{1,3}$ when convenient.}}
\textit{The Clifford algebra $\mathcal{C}\ell_{p,q}$, associated to the quadratic space $\mathbb{R}^{p,q}$, is the associative unital algebra such that}

\begin{enumerate}
\item[(i)]\label{it1} \textit{The Clifford application $\gamma:\mathbb{R}^{p,q} \to \mathcal{C}\ell_{p,q}$ is linear and satisfies}\footnote{In general, as a matter of simplicity, the Clifford application is omitted and the Clifford product is denoted by juxtaposition.}
$$\gamma (v)\gamma (u)+\gamma (u)\gamma (v)=2g({ v, u}), \qquad \forall\, { v,u} \in \mathbb{R}^{p,q};$$
\textit{
\item[(ii)] If $(\mathcal{Y},\gamma')$ is another associative unital algebra and an application $\gamma ':\mathbb{R}^{p,q} \to  \mathcal{Y}$ satisfies $\gamma ' (v)\gamma ' (u)+\gamma ' (u)\gamma ' (v)=2g({ v, u})$, thus there is an unique homomorphism $\phi: \mathcal{C}\ell_{p,q}  \to \mathcal{Y}$ such that $\gamma'=\phi \circ \gamma$.}
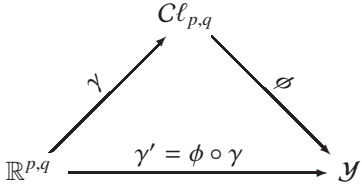
\begin{diagram}
 &  & \mathcal{C}\ell_{p,q} \ & &\\
 & \ruTo(2,2)^\gamma &  &  \rdTo(2,2)^{\phi} & \\
\mathbb{R}^{p,q}&  &\ \rTo^{ \gamma'=\phi \circ \gamma} &  & \mathcal{Y}
\end{diagram}
\end{enumerate}

Under the structure of Clifford algebras, besides the classical definition of spinors as elements of the space carrying the irreducible representation of the Spin group (the Lorentz group in the case of Minkowski space), there exists also the important, while less popular, algebraic definition \cite{chevalley1996algebraic}. Algebraic spinors are minimal left ideals built upon primitive idempotents of the base algebra. In general, given the Clifford algebra $\mathcal{C}\ell_{p,q}$ and $f$ a primitive idempotent, the minimal left ideals are of the form $\mathcal{C}\ell_{p,q}f$. Furthermore, a division ring $\mathbb{K}$ isomorphic to $\mathbb{R}, \mathbb{C}$ or $\mathbb{H}$ (reals, complex and quaternions) is obtained, depending on the dimension and signature  of the space, by $f\mathcal{C}\ell_{p,q}f$. The application
\begin{align}
\begin{array}{cccl}
\cdot&: \mathcal{C}\ell_{p,q}f \times \mathbb{K} & \rightarrow & \mathcal{C}\ell_{p,q}f  \\ 
 & (\psi,a) & \mapsto  & \psi \cdot a \equiv \psi a,
\end{array}
\end{align}
on the other hand, defines a right $\mathbb{K}$-module structure over $\mathcal{C}\ell_{p,q}f$. Provided with this right $\mathbb{K}$-module structure,  $\mathcal{C}\ell_{p,q}f$ is called \textit{algebraic spinor space} of $\mathcal{C}\ell_{p,q}$, denoted by  $\mathbb{S}_{p,q}$. Analogously, minimal right ideals can be built upon primitive idempotents, having the form $f\mathcal{C}\ell_{p,q}$. The above division ring and $\mathbb{K}$-module structure are also analogous for minimal right ideals. The algebraic spinor space of right ideals of the algebra $\mathbb{C}\ell_{p,q}$ is denoted by $\mathbb{S}^\star_{p,q}$. The action of an element of  $\mathbb{S}^\star_{p,q}$ on the left hand side of   $\mathbb{S}_{p,q}$ defines a linear application whose image is the division ring $\mathbb{K}$. Moreover,  $\mathbb{S}^\star_{p,q}$ is isomorphic to  $\mathcal{L}(\mathbb{S}_{p,q}, \mathbb{K})$, the space of the linear applications of  $\mathbb{S}_{p,q}$ on $\mathbb{K}$, motivating the investigation of inner products on algebraic spinors space. In fact, given $\mathbb{S}^\star_{p,q} \simeq \mathcal{L}(\mathbb{S}_{p,q}, \mathbb{K})$, an inner product $\beta: \mathbb{S}_{p,q} \times \mathbb{S}_{p,q} \to \mathbb{K}$ is defined by  associating an arbitrary spinor $\psi \in \mathbb{S}_{p,q}$ to its correspondent $\psi^\star \in \mathbb{S}^\star_{p,q}$, called the adjoint with respect to the inner product $\beta$, such that $\beta(\psi, \phi)=\psi^\star \phi \in \mathbb{K}$. 

Right ideals are mapped into left ideals, and vice-versa, by involutions of the algebra. Idempotents, however, are not always preserved. In other words, denoting a generic involution by $\alpha$, follows $\alpha(\mathcal{C}\ell_{p,q} f)=\alpha(f)\,\alpha(\mathcal{C}\ell_{pq})=\alpha(f)\,\mathcal{C}\ell_{pq}$. Nevertheless $\alpha(f) \neq f$ in general and consequently $\alpha(f)\,\mathcal{C}\ell_{pq} \neq f\,\mathcal{C}\ell_{pq}$. Notwithstanding, there always exists an $h\in \mathcal{C}\ell_{p,q}$ such that $\alpha(f)=h^{-1}f\,h$ and $\alpha(h)= h$ \cite{Vaz:2016qyw, Benn:1987fw}, allowing one to define $\psi^\star=\alpha(h\psi)=h\,\alpha(\psi)=h\,\alpha(\psi f)=f\, h\,\alpha(\psi)$. Thus, an inner product can be obtained by 
\begin{equation}
\beta(\psi, \phi)=h\,\alpha(\psi)\phi=f\, h\,\alpha(\psi)\phi f \in f\mathcal{C}\ell_{p,q}f \simeq \mathbb{K},
\end{equation}
where $\alpha$ is called the adjoint involution of the inner product $\beta$. There are two natural involutions inside the structure of Clifford algebras, the reversion and Clifford conjugation \cite{Vaz:2016qyw}, these two involutions determine two inner products on $\mathbb{S}_{p,q}$, being any other inner product on $\mathbb{S}_{p,q}$ determined by an equivalent involution\cite{Benn:1987fw}. 
When dealing with complexified Clifford algebras $\mathbb{C}\otimes \mathbb{C}\ell_{p,q}$, the composition of the complex conjugation with the other algebra involutions changes the adjoint involution of inner products. The important situation for us is the one of the adjoint involution being equivalent to the hermitian conjugation on the algebra representation. For such it is sufficient that  $\alpha^*(a)=h^{-1}a^\dagger h$ and $h^\dagger = h$, for any $a\in \mathbb{C}\otimes \mathcal{C}\ell_{p,q}$ (in particular for $a\in \mathbb{C}\otimes \mathbb{S}_{p,q}$) and $h\in \mathbb{C}\otimes \mathcal{C}\ell_{p,q}$ \cite{Vaz:2016qyw, Benn:1987fw}. The adjoint (dual) spinor $\psi^\star\in \mathbb{C}\otimes \mathbb{S}^\star_{p,q}$ thus reads
\begin{equation}\label{algdual}
\psi^\star=h\alpha^*(\psi)=\psi^\dagger h=[h\psi]^\dagger.
\end{equation}

At this point one can notice the similarity between the duals of equations \eqref{1} and \eqref{algdual}. We argue that they are actually the same. As can be easily checked, 
one has just to take $h=\eta\Delta$, with $\eta^\dagger=\eta$, in order to have $\tilde{\psi}=\psi^\star$. In addition, an important fact not mentioned in previous papers is that $\Delta$ must be such that 
\begin{equation}\label{delta}
\Delta^\dagger \eta=\eta \Delta.
\end{equation}
It imposes a restriction to the operator $\Delta$, apart from being invertible, as discussed below \footnote{Notice that, despite the standard spacetime algebra (or Dirac algebra) being $\mathbb{C}\otimes \mathcal{C}\ell_{1,3}$, with $\mathbb{K}\simeq \mathbb{C}$, the algebraic discussion of this section covers the general case.}. The particular case of the Dirac dual, with $\Delta=\mathbb{I}$, obviously satisfies \eqref{delta}. The alternative Dirac dual found in Ref. \cite{Rogerio:2017hwz}, which was shown to have identical observables,  also obey Eq. \eqref{delta}\footnote{From here to the end of this section we are going to take $\eta=\gamma^0$, which comes from Lorentz covariance, as justified at the end of the section.}. For the mass dimension one duals proposed in Ref. \cite{Ahluwalia:2016rwl}, besides the one of Eq. \eqref{1}, the duals of Eqs. \eqref{A} and \eqref{B} also satisfy ${h}^\dagger=h$, for $h=\mathcal{A}\eta\Delta$ or $h=\mathcal{B}\eta\Delta$. In fact, given $\mathcal{A}=2[\mathbb{I}-\tau G(\phi)]^{-1}$ (see Sec. \ref{auto} or Ref. \cite{Ahluwalia:2016rwl}), as $\mathcal{A}^\dagger=\mathcal{A}$ and the commutators $[\mathcal{A},\Delta]$ and $[\mathcal{A},\eta]$ do vanish, it is straightforward to check that $(\mathcal{A}\eta \Delta)^\dagger=\mathcal{A}\eta \Delta$, thus
\begin{equation}
\stackrel{\neg}{\lambda}\,\!\!^S= [\Delta {\lambda}\,\!\!^S]^{\dagger}\eta \mathcal{A}=[\mathcal{A}\eta \Delta {\lambda}\,\!\!^S]^{\dagger}=[ {\lambda}\,\!\!^S]^{\dagger} \mathcal{A}\eta \Delta.
\end{equation}
The same happens for $\mathcal{B}$. Notice that the above duals are equivalent to redefining $\Delta$ as $\stackrel{\neg}{\Delta}=\mathcal{A}{\Delta}$ or $\stackrel{\neg}{\Delta}=\mathcal{B}{\Delta}$, indicating the $\Delta$ operator as the one carrying the generality of the new duals. This is an important aspect. With effect, it is the mathematical counterpart to the alluded remain physical degree of indeterminacy in Ref. \cite{Ahluwalia:2016rwl}, responsible to make possible this first part of the new dual formulation.  

A further characterization of the $\Delta$ operator can be found by taking a general complex matrix $\Delta=[a_{ij}]$ and imposing the Eq. \eqref{delta}. It gives
\begin{equation}
\Delta=\left[\begin{array}{cccc}
a_{11} & a_{12} & a_{13} & a_{14} \\ 
a_{21} & a_{22} & a^*_{14} & a_{24} \\ 
a_{31} & a_{32} & a^*_{11} & a^*_{21} \\ 
a^*_{32} & a_{42} & a^*_{12} & a^*_{22}
\end{array}\right], \;\; \text{with}\;\; a_{13},a_{31},a_{24},a_{42} \in \mathbb{R}. 
\end{equation}
The above matrix has a clear structure of block matrix, which, as expected, is compatible with the one found in \cite{Ahluwalia:2016rwl} and in \cite{Rogerio:2017hwz}. It can be represented by
\begin{equation}
\Delta=\left[\begin{array}{cccc}
A & B  \\ 
C & A^\dagger 
\end{array}\right],\; \text{with}\; B^\dagger=B \;\;\text{and}\;\;  C^\dagger=C. 
\end{equation}
Explicitly, for the dual introduced in Ref. \cite{Ahluwalia:2016rwl}, given the four-momentum $p^\mu=(E, p \sin \theta \cos \phi, p \sin \theta \sin \phi, p \cos \theta)$, hold $B=C=0$ and 
\begin{equation}
A=\left[\begin{array}{cccc}
\frac{i p \sin \theta }{m} & -\frac{i e^{-i \phi } (E-p \cos \theta )}{m}  \\ 
\frac{i e^{i \phi } (E-p \cos \theta )}{m} & -\frac{i p \sin \theta }{m} 
\end{array}\right]. 
\end{equation}
For the one found in Ref. \cite{Rogerio:2017hwz}, on the other hand, hold $A=0$, 
\begin{equation}
B=\left[\begin{array}{cccc}
\frac{m}{E-p} & 0  \\ 
0 & \frac{m}{E-p}
\end{array}\right]\quad \mbox{and} \quad C=\left[\begin{array}{cccc}
\frac{m}{E+p} & 0  \\ 
0 & \frac{m}{E+p}
\end{array}\right]. 
\end{equation}
 
Now let us take a look at the Lorentz invariance of the inner product. Denoting by $D(\Lambda)$ the Lorentz transformation in the $(1/2,0)\oplus(0,1/2)$ representation space and taking $\psi'=D(\Lambda)\psi$ and $\Delta'=D(\Lambda)\Delta D(\Lambda)^{-1}$, the Lorentz invariance imposes  
\begin{equation}\label{loinv}
\tilde{\psi'}\psi'=\tilde{\psi}\psi.
\end{equation}
The left hand side of the above equation reads
\begin{equation}\label{dpsi}
\tilde{\psi'}\psi'={\psi'}^\dagger \Delta'^\dagger\eta\psi'={\psi}^\dagger \Delta^\dagger D(\Lambda)^\dagger\eta D(\Lambda)\psi.
\end{equation}
From Eqs. \eqref{loinv} and \eqref{dpsi} follows $\eta D(\Lambda)^\dagger \eta=D(\Lambda)^{-1}$, which is the well known relation found for Dirac duals that, along with the parity symmetry, fixates $\eta=\gamma^0$ \cite{Ahluwalia:2016rwl}. It means that the $\Delta$ operator does not affect the Lorentz invariance of the inner product, this task is fulfilled by $\eta$, evidencing again $\Delta$ as being the only responsible for the degrees of freedom of a general Lorentz covariant dual.

\section{The Automorphism structure and an attempt of interpretation}\label{auto}

In order to further analyze the second part of the new dual formulation, we shall start this section by depicting the relevant properties associated to the operators $\mathcal{A}$ and $\mathcal{B}$ of mass dimension one duals. It allows for the investigation of spinorial spaces underlying automorphisms, evincing the form of $\mathcal{A}$ (and consequently $\mathcal{B}$) by construction, followed by the study of the relationship between the so called $\tau$-deformation and symmetries of such spaces.  

Classes of spinors do not form, in general, vector spaces. However, as discussed in \cite{Cavalcanti:2015adn}, there are special cases where this task can be accomplished. Two of them are the space of self conjugate and anti-self conjugate Elko. Taking elements of those vectorial spaces, in such a way that $\lambda^S \in \mathbb{E}_S$ and $\tilde{\lambda}^S\in \mathbb{E}_S^*$, the action of $\mathbb{E}^*_S$ elements is taken with respect to elements of $\mathbb{E}_S$, i. e., $\tilde{\lambda}^S: \mathbb{E}_S\rightarrow \mathbb{C}$. Notice that by means of the anti-self conjugated space $\mathbb{E}_A \ni \lambda^A$, the action of $\mathbb{E}^*_S$ elements may be naturally extended to  $\tilde{\lambda}^S: \mathbb{E}_S\oplus\mathbb{E}_A\rightarrow \mathbb{C}$, since $\tilde{\lambda}^S\lambda^A=0$.  

According to Secs. \ref{prel} and \ref{algebraic}, in order to reach a Lorentz invariant spin sum, a redefinition of the dual elements takes place. Therefore, after defining 
\begin{equation}
\stackrel{\neg}{\lambda}\,\!\!^S=\tilde{\lambda}^S\mathcal{A},\label{re0}
\end{equation} it is important to set down the properties of the new operator $\mathcal{A}$. We shall pay attention to the $\mathcal{A}$ operator case, being the $\mathcal{B}$ case completely analog. In the course of this exposition we touch on the peculiarities of this last case only when necessary. 

The first property required for the new dual is 
\begin{equation}
\stackrel{\neg}{\lambda}\,\!\!^S\lambda^S=\tilde{\lambda}^S\lambda^S,\label{re1}
\end{equation} from which we have $\mathcal{A}\lambda^S=\lambda^S$. The remaining requirement regarding the action of $\mathcal{A}$ is inferred by imposing $\stackrel{\neg}{\lambda}\,\!\!^S\lambda^A=0$. Since $\tilde{\lambda}^S\lambda^A=0$, it is quite enough that $\mathcal{A}\lambda^A=\sigma\lambda^A$ for some non vanishing $\sigma \in\mathbb{C}$. These eigenspinors relations allows the action of this operator be such that 
\begin{equation}
\mathcal{A}:\mathbb{E}_S\oplus\mathbb{E}_A\rightarrow \mathbb{E}_S\oplus\mathbb{E}_A, \label{re2}
\end{equation} linearly. Therefore the domain $\mathcal{D}$ and range of $\mathcal{A}$ both coincide with the whole vectorial space: $\mathcal{D}(\mathcal{A})=\mathbb{E}_S\oplus\mathbb{E}_A=ran(\mathcal{A})$. Notice that the new dual share the main properties of the old dual 
\begin{eqnarray}
\stackrel{\neg}{\lambda}\,\!\!^S \in \mathbb{E}_S^*,\label{re3}\\ \mathcal{D}(\stackrel{\neg}{\lambda}\,\!\!^S)=\mathbb{E}_S\oplus\mathbb{E}_A=\mathcal{D}(\tilde{\lambda}^S),\label{re4} \\ ran(\stackrel{\neg}{\lambda}\,\!\!^S)=\mathbb{C}=ran(\tilde{\lambda}^S). \label{re5} 
\end{eqnarray}
Moreover, it is fairly simple to see that $ker(\stackrel{\neg}{\lambda}\,\!\!^S)=\mathbb{E}_A=ker(\tilde{\lambda}^S)$, and therefore the kernel associated to $\mathcal{A}$ is a bit more constrained. In fact, suppose a given spinor, say $\chi$, of $\mathbb{E}_S\oplus\mathbb{E}_A$ belonging to $ker(\mathcal{A})$. Then $\mathcal{A}\chi=0$ and taking into account the evinced action of $\mathcal{A}$ on the basis of the space,  $\mathcal{A}\chi=0$ implies $\chi=0$ necessarily. Therefore $ker(\mathcal{A})=\{0\}$ and $\mathcal{A}$ is an injective operator. In addition, for a general spinor $\lambda=\alpha \lambda^S+\beta\lambda^A \in\mathbb{E}_S\oplus\mathbb{E}_A$, there is always another spinor $\lambda_0=\alpha \lambda^S+\frac{\beta}{\sigma}\lambda^A \in \mathbb{E}_S\oplus\mathbb{E}_A$ such that $\mathcal{A}\lambda_0=\lambda$, and hence $\mathcal{A}$ is surjective. Combining all these properties, we see that the $\mathcal{A}$ operator engenders an isomorphism of the space in itself. 

Usually, the elements falling into this classification belongs to the automorphism set. Nevertheless, we shall pinpoint that a sufficient requirement to additionally justify the $\tau-$deformation is to impose $\mathcal{A}$ is a member of the automorphism group, i. e. $\mathcal{A}\in Aut(\mathbb{E}_S\oplus\mathbb{E}_A)$. Notice that it does not fixate the operator $\mathcal{A}$. The charge conjugation operator, for instance, naturally share with $\mathcal{A}$ all the above properties. The idea behind the introduction of the new dual is to achieve a Lorentz invariant spin sum, therefore $\mathcal{A}$ must serve (also) as an inverse to $(1+G(\phi))$. It turns out, however, that $(1+G(\phi))$ is not an element of $Aut(\mathbb{E}_S\oplus\mathbb{E}_A)$. In fact $(1+G(\phi))\lambda^A=0$ for all $\lambda^A \in \mathbb{E}_A$ and therefore $ker(1+G(\phi))=\mathbb{E}_A$, resulting in a non injective operator. Despite $(1+G(\phi))$ being not an element of the automorphism group, its $\tau-$deformation $(1+\tau G(\phi))$, with $\tau \in \mathbb{R}$, belongs to $Aut(\mathbb{E}_S\oplus\mathbb{E}_A)$. In fact, as it can be readily verified, $(1+\tau G(\phi))\lambda^S=(1+\tau)\lambda^S$ and $(1+\tau G(\phi))\lambda^A=(1-\tau)\lambda^A$, leading to $ker(1+\tau G(\phi))=\{0\}$. As elements of the same group, it is possible to impose $\mathcal{A}$ as the inverse of $(1+\tau G(\phi))$. Also, as it is evident from the construction, one must undertake the $\tau\rightarrow 1$ limit at the end of the calculations \cite{Lee:2014opa}. As shown in Ref. \cite{Rogerio:2016mxi}, this limit is formal in linear algebra. The reasoning exposed here demonstrate that the very nature of the $\mathcal{A}$ operator, better saying, the necessary properties of $\mathcal{A}$, make it a possible element of the automorphism group, and hence, in order to act as an inverse, the $\tau-$deformation is indeed necessary. 

Now, with a suitable normalization constant, the $\mathcal{A}$ operator may be written as \cite{Lee:2014opa, Ahluwalia:2016rwl}
\begin{equation}
\mathcal{A}=2\Bigg(\frac{\mathbb{I}+\tau G(\phi)}{1-\tau^2}\Bigg)_{\tau\rightarrow 1}\tilde{A}, \label{quase2}
\end{equation} with $\tilde{A}$ Lorentz invariant. Note that all the previous construction seems not to fix completely the form of $\mathcal{A}$. In fact, if the criteria for the spin sums is Lorentz invariance, then at first sight it is still possible to have an element of arbitrariness encoded in $\tilde{A}$. Let us deal with this issue now. Obviously, it is necessary that $\tilde{A}\in Aut(\mathbb{E}_S\oplus\mathbb{E}_A)$ and $\tilde{A}\lambda^S=\lambda^S$. Besides $\tilde{A}$ must be dimensionless. While these requirements do constraint the functional form of $\tilde{A}$, the reason why $\tilde{A}$ must be identified to the identity comes not from the mathematical framework, but is a physical requirement. The spin sums are at the very heart of quantum field theory. Entering in the propagator, the spin sums cascade down to several relevant quantum process and quantum interpretation of the theory as well. Being $\tilde{A}$ different from the identity, the dynamics of the associated quantum field would be completely different from every known dynamics. As the usual Dirac dynamics is automatically excluded (the spinors at hand are not annihilated by Dirac operator), the only possibility is $\tilde{A}=\mathbb{I}$. At the end, as the inverse part of the operator is unique, the functional form of $\mathcal{A}$ is obtained in an exhaustive fashion. Finally, the output of this construction is the spin sums equal to $\pm m \mathbb{I}$ (or proportional to $e_{Aut(\mathbb{E}_S\oplus\mathbb{E}_A)}$, the automorphism group identity) which is trivially Lorentz invariant.

We would like to make the parenthetical remark that the spin sums for the usual Dirac case also show to exhibit a result not belonging to the automorphism group. The crucial, and obvious, difference is that in the Dirac case the spin sums are automatically Lorentz invariant. 

The precedent formalization may also be accompanied of an attempt to interpret the effect leaded by the $\tau-$deformation. As a matter of fact, it was shown that the kernel of $(1\pm \tau G(\phi))$ jumps to $\mathbb{E}_{A/S}$ in the limit $\tau\rightarrow 1$. Besides, as it is clear from the previous procedure, we depart from $HOM(2)$ or $SIM(2)$ symmetries to the full Lorentz group. However, differently from usual spacetime symmetries breaking in which one has the emergence of an order parameter dictating the breaking effects magnitude, the relationship of the full Lorentz group and its relativistic avatars is not mediated by local symmetry breaking operators. The eventual interchange between these symmetries is not performed smoothly, in a manner of speaking. In order to give a mathematical sense at the symmetry level to the $\tau-$deformation, we start by defining $X$ as the set whose elements are the spin sums before the deformation. Besides, let $H$ be a group and consider a particular group action given by 
\begin{eqnarray}
\triangleright : H\times X\rightarrow X \nonumber \\ (h\in H, x\in X)\mapsto h \triangleright x=x,\label{action}
\end{eqnarray} for all $x$ of $X$ with $h$ different from $e_H$. There is a particular suitable realization for this action for $H=HOM(2)\ni \Gamma$ and 
\begin{eqnarray}
\Gamma\triangleright x:= \Gamma x\Gamma^{-1}=x. \label{rere}
\end{eqnarray} Notice that this realization is in fact that one performed in Ref. \cite{Ahluwalia:2008xi} acting on $G(\phi)$ and entailing a VSR symmetry to it and, obviously, to the spin sums. With the aid of (\ref{rere}), we have as the orbit of $x$ the following set
\begin{eqnarray}
\mathcal{O}(x)=\{\Gamma\triangleright x= \Gamma x\Gamma^{-1}=x| \Gamma\in HOM(2)\}. \label{orb}
\end{eqnarray}

Now consider, in analogous fashion, the set $X_\tau$ of spin sums after the tau-deformation and the recognition of $\mathcal{A}$ and $\mathcal{B}$ as the suitable inverses in the previously discussed context. Consider also a given group $G$ and a action group defined as in (\ref{action}) with $H$ and $X$ replaced by $G$ and $X_\tau$, respectively. Take $G=SO(3,1)$ plus discrete symmetries, the full Lorentz group. While it is obvious that the elements $\Gamma \in HOM(2)$ can act on $X_\tau$ (notice that the residual element of the spin sums contained in $X_{\tau}$ is the rest mass, which is obviously invariant for both groups), the opposite is no longer true. Thus elements $\Lambda \in G$ cannot act on $X$ by means of (\ref{action}) and the orbit (\ref{orb}) is more restricted then the orbit associated to $x_\tau \in X_\tau$. In fact
\begin{eqnarray}
\mathcal{O}(x_\tau)=\{\Gamma\triangleright x\,|\,\Gamma\in HOM(2)\}\cup\{\Lambda\triangleright x_\tau \,|\, \Lambda \in G\},\label{orb2}
\end{eqnarray} and therefore $\mathcal{O}(x_\tau)\supset \mathcal{O}(x)$. This means that the $\tau-$deformation and the new dual have enlarged the orbit of the spin sums. This is precisely what is expected from a symmetry interchange which is not performed by means of a local operator. Besides, even considering different approaches to the symmetry breaking phenomena \cite{Frohlich:1981yi}, the orbits are expected to be disjoint and addressed to the same set, whilst here we have partially overlapping orbits related to different sets. We shall interpret it as a mathematical evidence against the spontaneous breaking of $G$ down to $HOM(2)$ (or vice-versa in the case at hand). This fact, we believe, gives support to the formulation we approached here to the spin sums, i. e., the physical requirement of Lorentz invariant spin sums must be imposed in the elaboration of the theoretical framework, as it is not expected a dynamical (spontaneous) mechanism engendering the transition from $HOM(2)$ to the full Lorentz group.

\section{Final Remarks}

In this paper we investigate some additional formal features of the Lorentz preserving mass dimension one duals. The non standard new dual was shown to be completely consistent with the mathematically well established duals of algebraic spinors, from Clifford algebras. Moreover, we shown that the general form of the dual proposed in \cite{Ahluwalia:2016rwl}, obeying an algebraic constrain, is actually the most general Lorentz preserving dual. It makes clear that all its the freedom of choice is encapsulated by the  $\Delta$ operator. The algebraic constrain of Eq. \eqref{delta} allowed us to find a general form for the $\Delta$ operator in its matrix representation.  The preceding results are followed by a though analysis of the hole of the $\tau$-deformation in a hidden automorphism structure of the self conjugate and anti-self conjugate spinors space as well as in orbits of such spaces under the action of VSR groups. We argue that, combining our results with the ones obtained in Refs. \cite{Ahluwalia:2016rwl, Rogerio:2016mxi} and \cite{Rogerio:2017hwz}, it is possible to claim that mass dimension one dual theory as not only physically but also mathematically consistent.

\section{Acknowledgments}
JMHS thanks to CNPq for partial support. RCT thanks the UNESP-Guaratinguet\'a Post-Graduation program and CAPES. 



\bibliographystyle{elsarticle-num} 
\bibliography{dual.bib}

\begin{thebibliography}{10}
\expandafter\ifx\csname url\endcsname\relax
  \def\url#1{\texttt{#1}}\fi
\expandafter\ifx\csname urlprefix\endcsname\relax\def\urlprefix{URL }\fi
\expandafter\ifx\csname href\endcsname\relax
  \def\href#1#2{#2} \def\path#1{#1}\fi

\bibitem{Ahluwalia:2004sz}
D.~V. Ahluwalia, D.~Grumiller, {Dark matter: A Spin one half fermion field with
  mass dimension one?}, Phys. Rev. D 72 (2005) 067701.
\newblock \href {http://arxiv.org/abs/hep-th/0410192}
  {\path{arXiv:hep-th/0410192}}, \href
  {https://doi.org/10.1103/PhysRevD.72.067701}
  {\path{doi:10.1103/PhysRevD.72.067701}}.

\bibitem{Ahluwalia:2004ab}
D.~V. Ahluwalia, D.~Grumiller, {Spin half fermions with mass dimension one:
  Theory, phenomenology, and dark matter}, JCAP 0507 (2005) 012.
\newblock \href {http://arxiv.org/abs/hep-th/0412080}
  {\path{arXiv:hep-th/0412080}}, \href
  {https://doi.org/10.1088/1475-7516/2005/07/012}
  {\path{doi:10.1088/1475-7516/2005/07/012}}.

\bibitem{daRocha:2011yr}
R.~da~Rocha, A.~E. Bernardini, J.~M. Hoff~da Silva, {Exotic Dark Spinor
  Fields}, JHEP 04 (2011) 110.
\newblock \href {http://arxiv.org/abs/1103.4759} {\path{arXiv:1103.4759}},
  \href {https://doi.org/10.1007/JHEP04(2011)110}
  {\path{doi:10.1007/JHEP04(2011)110}}.

\bibitem{Lounesto:2001zz}
P.~Lounesto, Clifford algebras and spinors, Vol. 286 of Lond. Math. Soc. Lect.
  Note Ser., Cambridge University Press, 2001.

\bibitem{Cavalcanti:2014wia}
R.~T. Cavalcanti, {Classification of Singular Spinor Fields and Other Mass
  Dimension One Fermions}, Int. J. Mod. Phys. D 23~(14) (2014) 1444002.
\newblock \href {http://arxiv.org/abs/1408.0720} {\path{arXiv:1408.0720}},
  \href {https://doi.org/10.1142/S0218271814440027}
  {\path{doi:10.1142/S0218271814440027}}.

\bibitem{HoffdaSilva:2017waf}
J.~M. Hoff~da Silva, R.~T. Cavalcanti, {Revealing how different spinors can be:
  the Lounesto spinor classification}, Mod. Phys. Lett. A32~(35) (2017)
  1730032.
\newblock \href {http://arxiv.org/abs/1708.06222} {\path{arXiv:1708.06222}},
  \href {https://doi.org/10.1142/S0217732317300324}
  {\path{doi:10.1142/S0217732317300324}}.

\bibitem{Fabbri:2017lvu}
L.~Fabbri, R.~da~Rocha, {Unveiling a spinor field classification with
  non-Abelian gauge symmetries}, Phys. Lett. B780 (2018) 427--431.
\newblock \href {http://arxiv.org/abs/1711.07873} {\path{arXiv:1711.07873}},
  \href {https://doi.org/10.1016/j.physletb.2018.03.029}
  {\path{doi:10.1016/j.physletb.2018.03.029}}.

\bibitem{Lee:2018bfv}
C.-Y. Lee, {Mass dimension one fermions from flag dipole spinors}\href
  {http://arxiv.org/abs/1809.04381} {\path{arXiv:1809.04381}}.

\bibitem{daRocha:2013qhu}
R.~da~Rocha, L.~Fabbri, J.~M. Hoff~da Silva, R.~T. Cavalcanti, J.~A.
  Silva-Neto, {Flag-Dipole Spinor Fields in ESK Gravities}, J. Math. Phys. 54
  (2013) 102505.
\newblock \href {http://arxiv.org/abs/1302.2262} {\path{arXiv:1302.2262}},
  \href {https://doi.org/10.1063/1.4826499} {\path{doi:10.1063/1.4826499}}.

\bibitem{daRocha:2016bil}
R.~da~Rocha, R.~T. Cavalcanti, {Flag-dipole and flagpole spinor fluid flows in
  Kerr spacetimes}, Phys. Atom. Nucl. 80~(2) (2017) 329--333.
\newblock \href {http://arxiv.org/abs/1602.02441} {\path{arXiv:1602.02441}},
  \href {https://doi.org/10.1134/S1063778817020235}
  {\path{doi:10.1134/S1063778817020235}}.

\bibitem{deOliveira:2013hda}
E.~Capelas~de Oliveira, W.~A. Rodrigues, Jr., J.~Vaz, Jr., {Elko Spinor Fields
  and Massive Magnetic Like Monopoles}, Int. J. Theor. Phys. 53~(12) (2014)
  4381--4401.
\newblock \href {http://arxiv.org/abs/1306.4645} {\path{arXiv:1306.4645}},
  \href {https://doi.org/10.1007/s10773-014-2188-4}
  {\path{doi:10.1007/s10773-014-2188-4}}.

\bibitem{Fabbri:2014foa}
L.~Fabbri, S.~Vignolo, {ELKO and Dirac Spinors seen from Torsion}, Int. J. Mod.
  Phys. D23~(14) (2014) 1444001.
\newblock \href {http://arxiv.org/abs/1407.8237} {\path{arXiv:1407.8237}},
  \href {https://doi.org/10.1142/S0218271814440015}
  {\path{doi:10.1142/S0218271814440015}}.

\bibitem{Vaz:2017fac}
J.~Vaz, {The Clifford Algebra of Physical Space and Elko Spinors}, Int. J.
  Theor. Phys. 57~(2) (2018) 582--601.
\newblock \href {https://doi.org/10.1007/s10773-017-3591-4}
  {\path{doi:10.1007/s10773-017-3591-4}}.

\bibitem{Fabbri:2017pwp}
L.~Fabbri, {General Dynamics of Spinors}, Adv. Appl. Clifford Algebras 27~(4)
  (2017) 2901--2920.
\newblock \href {http://arxiv.org/abs/1707.03270} {\path{arXiv:1707.03270}},
  \href {https://doi.org/10.1007/s00006-017-0816-9}
  {\path{doi:10.1007/s00006-017-0816-9}}.

\bibitem{Ahluwalia:2016rwl}
D.~V. Ahluwalia, {The theory of local mass dimension one fermions of spin one
  half}, Adv. Appl. Clifford Algebras 27~(3) (2017) 2247--2285.
\newblock \href {http://arxiv.org/abs/1601.03188} {\path{arXiv:1601.03188}},
  \href {https://doi.org/10.1007/s00006-017-0775-1}
  {\path{doi:10.1007/s00006-017-0775-1}}.

\bibitem{Benn:1987fw}
I.~M. Benn, R.~W. Tucker, An Introduction to Spinors and Geometry With
  Applications in Physics, Adam Hilger LTD, 1987.

\bibitem{Rogerio:2016mxi}
R.~J.~B. Rog\'erio, J.~M. Hoff~da Silva, {The local vicinity of spins sum for
  certain mass dimension one spinors}, EPL 118~(1) (2017) 10003.
\newblock \href {http://arxiv.org/abs/1602.05871} {\path{arXiv:1602.05871}},
  \href {https://doi.org/10.1209/0295-5075/118/10003}
  {\path{doi:10.1209/0295-5075/118/10003}}.

\bibitem{Cohen:2006ky}
A.~G. Cohen, S.~L. Glashow, {Very special relativity}, Phys. Rev. Lett. 97
  (2006) 021601.
\newblock \href {http://arxiv.org/abs/hep-ph/0601236}
  {\path{arXiv:hep-ph/0601236}}, \href
  {https://doi.org/10.1103/PhysRevLett.97.021601}
  {\path{doi:10.1103/PhysRevLett.97.021601}}.

\bibitem{Lee:2015jpa}
C.-Y. Lee, {Symmetries and unitary interactions of mass dimension one fermionic
  dark matter}, Int. J. Mod. Phys. A31~(35) (2016) 1650187.
\newblock \href {http://arxiv.org/abs/1510.04983} {\path{arXiv:1510.04983}},
  \href {https://doi.org/10.1142/S0217751X16501876}
  {\path{doi:10.1142/S0217751X16501876}}.

\bibitem{Lee:2015sqj}
C.-Y. Lee, M.~Dias, {Constraints on mass dimension one fermionic dark matter
  from the Yukawa interaction}, Phys. Rev. D94~(6) (2016) 065020.
\newblock \href {http://arxiv.org/abs/1511.01160} {\path{arXiv:1511.01160}},
  \href {https://doi.org/10.1103/PhysRevD.94.065020}
  {\path{doi:10.1103/PhysRevD.94.065020}}.

\bibitem{Ahluwalia:2010zn}
D.~V. Ahluwalia, S.~P. Horvath, {Very special relativity as relativity of dark
  matter: The Elko connection}, JHEP 11 (2010) 078.
\newblock \href {http://arxiv.org/abs/1008.0436} {\path{arXiv:1008.0436}},
  \href {https://doi.org/10.1007/JHEP11(2010)078}
  {\path{doi:10.1007/JHEP11(2010)078}}.

\bibitem{Ahluwalia:2008xi}
D.~V. Ahluwalia, C.-Y. Lee, D.~Schritt, {Elko as self-interacting fermionic
  dark matter with axis of locality}, Phys. Lett. B 687 (2010) 248--252.
\newblock \href {http://arxiv.org/abs/0804.1854} {\path{arXiv:0804.1854}},
  \href {https://doi.org/10.1016/j.physletb.2010.03.010}
  {\path{doi:10.1016/j.physletb.2010.03.010}}.

\bibitem{Lee:2014opa}
C.-Y. Lee, {A Lagrangian for mass dimension one fermionic dark matter}, Phys.
  Lett. B760 (2016) 164--169.
\newblock \href {http://arxiv.org/abs/1404.5307} {\path{arXiv:1404.5307}},
  \href {https://doi.org/10.1016/j.physletb.2016.06.064}
  {\path{doi:10.1016/j.physletb.2016.06.064}}.

\bibitem{Hitzer2013}
E.~Hitzer, T.~Nitta, Y.~Kuroe,
  \href{https://doi.org/10.1007/s00006-013-0378-4}{Applications of clifford's
  geometric algebra}, Adv. Appl. Clifford Algebras 23~(2) (2013) 377--404.
\newblock \href {https://doi.org/10.1007/s00006-013-0378-4}
  {\path{doi:10.1007/s00006-013-0378-4}}.
\newline\urlprefix\url{https://doi.org/10.1007/s00006-013-0378-4}

\bibitem{Vaz:2016qyw}
J.~Vaz, Jr., R.~da~Rocha, {An Introduction to Clifford Algebras and Spinors},
  Oxford University Press, 2016.

\bibitem{chevalley1996algebraic}
C.~Chevalley, The Algebraic Theory of Spinors and Clifford Algebras: Collected
  Works, Vol.~2, Springer Science \& Business Media, 1996.

\bibitem{Rogerio:2017hwz}
R.~J. Bueno~Rogerio, C.~H. Coronado~Villalobos, {Non-standard Dirac adjoint
  spinor: The emergence of a new dual}, EPL 121~(2) (2018) 21001.
\newblock \href {http://arxiv.org/abs/1711.07856} {\path{arXiv:1711.07856}},
  \href {https://doi.org/10.1209/0295-5075/121/21001}
  {\path{doi:10.1209/0295-5075/121/21001}}.

\bibitem{Cavalcanti:2015adn}
R.~T. Cavalcanti, R.~Rocha, J.~M. Hoff~da Silva, {Could Elko Spinor Fields
  Induce VSR Symmetry in the DKP (Meson) Algebra?}, Adv. Appl. Clifford
  Algebras 27~(1) (2017) 267--277.
\newblock \href {https://doi.org/10.1007/s00006-015-0563-8}
  {\path{doi:10.1007/s00006-015-0563-8}}.

\bibitem{Frohlich:1981yi}
J.~Frohlich, G.~Morchio, F.~Strocchi, Higgs phenomenon without symmetry
  breaking order parameter, Nucl. Phys. B190 (1981) 553--582.
\newblock \href {https://doi.org/10.1016/0550-3213(81)90448-X}
  {\path{doi:10.1016/0550-3213(81)90448-X}}.

\end{thebibliography}


%

\end{document}